**Machine learning–enabled design of sustainable alloys considering real-world constraints**

***(preprint, Oct 2026, Nature Sustainability)***

Quan Li [1, a, b], Zhilong Wang [1, c, d], Lei Shi [c, d], Baojing Feng [a], Changhui Liu [a], Jie Wang [a], Jingya Wang [a], Weisen Zheng [e], Matic Jovičević-Klug [b], Ziyuan Rao [a, b] *, Jinjin Li [c,] *, Dierk Raabe [b,] *, Xiaoqin Zeng [a,] *

[a] *National Engineering Research Center of Light Alloy Net Forming, School of Materials Science and Engineering, Shanghai Jiao Tong University, Shanghai, 200240, China*
[b] *Max Planck Institute for Sustainable Materials, Düsseldorf, 40237, Germany*
[c] *National Key Laboratory of Advanced Micro and Nano Manufacture Technology, Shanghai Jiao Tong University, Shanghai, 200240, China*
[d] *School of Integrated Circuits (School of Information Science and Electronic Engineering), Shanghai Jiao Tong University, Shanghai, 200240, China*
[e] *State Key Laboratory of Advanced Special Steels, School of Materials Science and Engineering, Shanghai University, Shanghai 200444, China*
[1] These authors contributed equally to this work.
[*] Correspondence to: Xiaoqin Zeng (xqzeng@sjtu.edu.cn), Jinjin Li (lijinjin@sjtu.edu.cn), Ziyuan Rao (ziyuanrao@sjtu.edu.cn ), and Dierk Raabe (d.raabe@mpi-susmat.de).

## Abstract

Designing sustainable alloys must reconcile high performance with real-world constraints, including feedstock scarcity and variability, processing and energy requirements, regulatory thresholds, and economic limitations. Addressing these constraints reaches far beyond today's conventional performance-focused frameworks. It needs to consider not only complex property–processing–performance trade-offs but also sustainable production pathways. Here, we propose a physics-informed ensemble (PIE) learning framework that integrates fundamental structure-property relationships with processing routes and sustainability criteria for mechanism-based alloy design. It predicts properties from composition–microstructure relationships and infers processing routes under sustainability constraints, establishing an interdisciplinary approach to identifying sustainable alloy solutions. We validate the approach for Mg alloys, a lightweight yet energy-intensive metallic material class, considering performance, cost, and carbon emissions. Trained on a benchmark dataset, the PIE model systematically explores plausible alloy compositions and microstructures to construct a digital library of millions of candidate alloys. From this space, we identify two sustainable Mg alloy solutions that overcome the conventional performance–cost–emission trade-off: a rare-earth-containing alloy increases yield and tensile strength by 72.8% and 61.1%, respectively, over alloys from the same family at comparable cost and emissions; and a rare-earth-free alloy reduces cost and carbon emissions by 40.4% and 88.9%, respectively, relative to its primary-feedstock counterpart, without compromising mechanical performance. This work introduces a general, physics-informed, and quantifiable framework for designing sustainable alloys under real-world constraints.

High-performance alloys enable and advance products and infrastructures in practically all industrial sectors [1]. Alloy development has traditionally focused on the enhancement of key functional and mechanical properties—such as strength, ductility, magnetization, and electrical conductivity—while ensuring their stability under harsh service and operating conditions. Today, global industries—including the metallurgical sector—are increasingly pursuing sustainability and circularity to mitigate accelerating climate change, prevent irreversible environmental damage, and address resource depletion as well as the growing geopolitical and economic instabilities associated with critical raw materials [2]. Accordingly, next-generation alloys must meet an extended set of metrics, encompassing not only mechanical and functional features but also low-energy production, economic viability (e.g., through scrap utilization or inexpensive alloying elements), and drastically reduced environmental impact [3].

These multi-objective requirements confront us with two fundamental challenges. First, key mechanisms behind desired alloy features extend across vast compositional, microstructural and processing design spaces and are often determined by kinetic transients rather than by thermodynamic bounds alone. They operate over multiple length scales, from atomic defects to macroscopic transport phenomena as well as their interactions. Second, the coexistence of multiple strengthening mechanisms and damage modes further complicates predictive design. As a result, traditional material design approaches have so far explored the practically infinite alloy design space only very sparsely (**Fig. 1a**), with only about several ~$10^3$ alloys created in laboratories.

Moreover, only a tiny fraction of these laboratory-discovered alloys transition to practical use and ultimately become commercial alloys–that is, alloys produced at an industrial scale, certified for engineering use, and circulated in regional or global markets. Most laboratory-discovered alloys never reach commercialization because they fail to meet critical real-world constraints—including production costs, manufacturing scalability, resource availability, energy consumption, recyclability, carbon emissions, and customer needs. These constraints that reach far beyond conventional property metrics are essential yet often underrated, and they ultimately determine whether a high-performance alloy can compete when it comes to practical adoption under real-world manufacturing and application conditions. The design of sustainable alloys is therefore inherently cross-disciplinary, requiring expertise across materials science, economics, and environmental domains (**Fig. 1b**) [3,4]. As these real-world metrics fall outside conventional alloy databases, information screening based on previously explored alloys is often limited and lacks translational impact data. Meeting these challenges, therefore, demands a novel systematic and predictive framework that combines fundamental physical principles with real-world production constraints and practical processing implementation.

Machine learning (ML) has emerged as a powerful tool for navigating high-dimensional alloy design spaces by learning correlative surrogate models from data [5-7]. In some cases, purely data-driven approaches uncover complex relationships directly from multivariate data without predefined physical assumptions [8]. To enhance physical interpretability, previous physics-informed ML approaches incorporate physics-based descriptors and mechanistic relationships to capture nonlinear relationships among composition, microstructure, processing, and properties [9-12]. However, identifying near-causal descriptors while balancing accuracy, model complexity, and computational cost remains key challenge [13-15]. More recently, sustainability-aware approaches have broadened alloy design beyond performance by screening candidate compositions against sustainability metrics [16-18]. These real-world considerations introduce additional layers of complexity that are difficult to embed in classical performance-focused frameworks [19-21]. Yet, accounting for them are essential, as they finally decide if a new material will find its market or not.

We therefore propose an interdisciplinary, end-to-end solution that integrates physical composition-processing-microstructure-property relationships with quantitative sustainability assessments, linking laboratory-scale alloy design with real-world alloy implementation, as shown in **Fig. 1c**. The workflow includes three main steps: (1) learning from laboratory data to design target alloys and establish their composition-microstructure-property (CMP) profile; (2) extending these CMP profiles to composition-microstructure-property-processing (CMPP) profiles by incorporating corresponding practical manufacturing-related metrics; and (3) performing further interdisciplinary optimization using the CMPP profiles in conjunction with quantitative sustainability assessments. This framework enables the design of sustainable alloys by expanding laboratory-scale data into a vast, unexplored, industry-oriented, and physics-based CMPP design space that far exceeds traditional and often highly idealized material exploration approaches.

Here, we focus on load-bearing alloys to demonstrate the potential of this framework. Air traffic, space vehicles, high-speed trains, and electric cars, which form the backbone of modern transportation, face intense pressure for weight reduction to improve energy efficiency (e.g., ~2% annual increase in aviation [22]) and reduce emissions in line with global net-zero $CO_2$ goals by 2050 [23]. To meet these needs, lightweight alloy design must advance along two directions [24]: (1) Indirect sustainability, in which alloys serve through superior performance, lower weight, and higher life times. (2) Direct sustainability, in which the production of alloys and parts causes lower environmental impact and improves recyclability while maintaining economic viability and performance.

As a model case, we chose here the specific case of magnesium (Mg), the lightest structural alloy, which possesses a high strength-to-weight ratio, exceeding that of Al by ~17% and that of iron by ~210% [25]. The use of Mg in automotive components enables a 22 – 70% weight reduction relative to Al alloys and steel, translating to an estimated 12 ~ 56% improvement in fuel economy [26,27]. However, Mg alloys in practice face critical challenges in production, sustainability, and cost due to their limited ductility and sheet formability, high chemical reactivity, strong affinity for oxygen, and extremely low reduction potential ($E^{o}$ = –2.37 V vs. SHE [28]). The production of Mg is therefore highly energy-intensive (~200 MJ/kg, ~10 times that of Fe [29]), leading to substantial carbon emissions (~23 kg $CO_2$-eq/kg, ~15 times that of Fe [30]), and high operational expenditures (~2400 EUR/ton). In addition, conventional primary production and downstream processing suffer from severe oxidation, melt loss, and high rejection rates, resulting in significant material losses of up to ~70% of the initial Mg mass before reaching the final product [31,32]. This low material utilization further enhances production costs and undermines Mg's sustainability benefits. Given these challenges, the practical imperative is to identify manufacturable Mg alloy solutions that meet high-performance requirements for end-use applications while satisfying sustainability goals for cost-effectiveness and emission compliance.

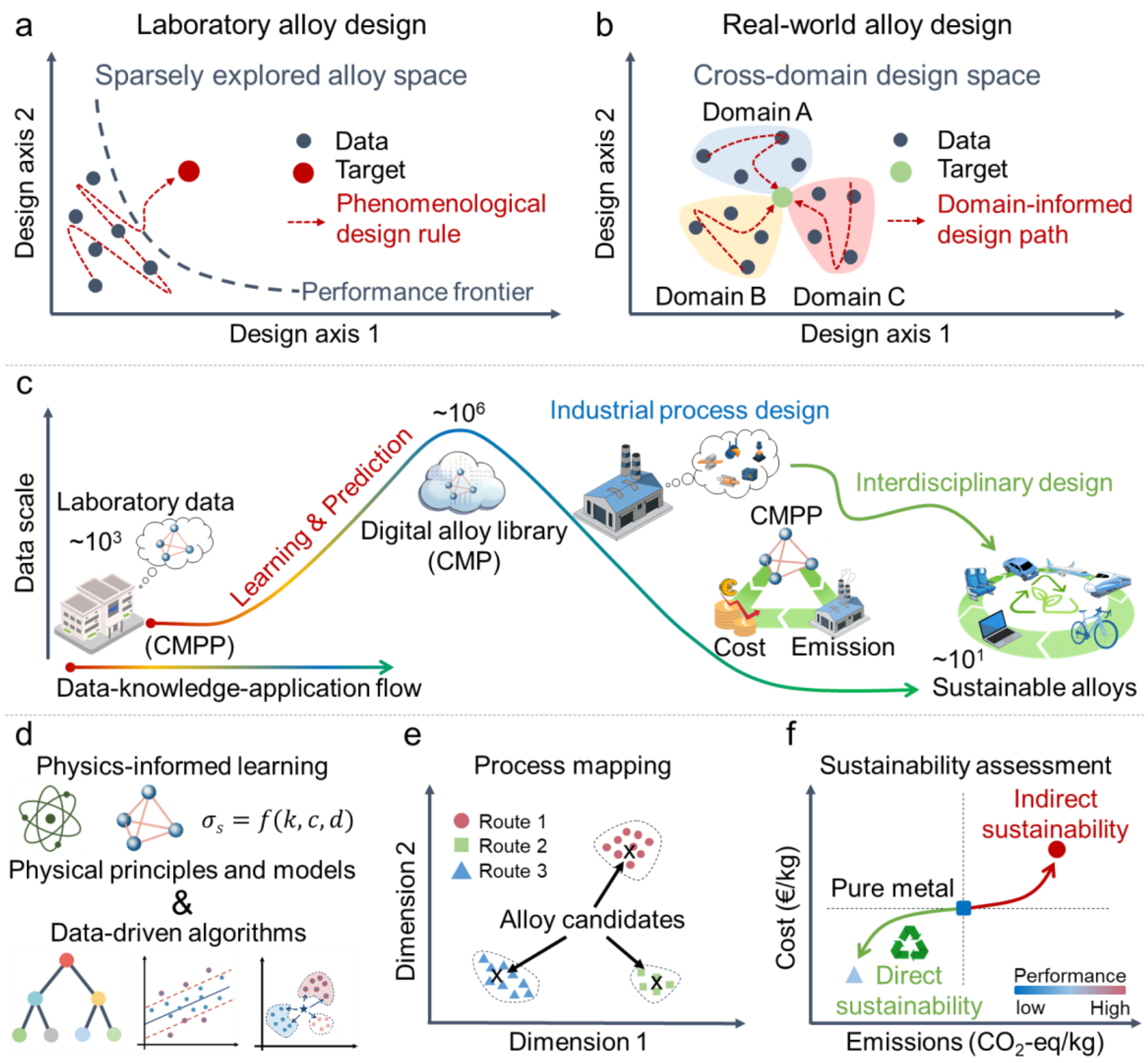

**Fig. 1 | Physics-informed machine learning workflow for sustainable alloy design. a.** Laboratory alloy design follows traditional empirical rules and relies on sparsely sampled experiments to explore alloys approaching or exceeding the performance frontier, typically focusing on performance trade-offs (e.g., strength vs. ductility). **b.** Real-world alloy design integrates multiple domains under coupled constraints, extending the design space from physical properties to multi-domain metrics. **c.** Conceptual data–knowledge–application workflow: a laboratory-scale dataset (e.g., ~$10^3$ entries per alloy system) is expanded into a digital alloy library (~$10^6$ candidates), followed by industrial processing-route identification and interdisciplinary optimization to identify a small set of sustainable alloy candidates (e.g., ~$10^1$). **d.** The physics-informed ensemble learning model integrates

multiscale physical principles, simplified physics-based models, and data-driven algorithms. **e**. Data-driven process mapping clusters candidate alloys with explored alloys to infer feasible processing routes. **f**. Application-oriented optimization balances performance, cost, and carbon emissions to identify (1) alloys that win through indirect sustainability (low weight, high performance), and (2) alloys that win through direct sustainability (lower emissions at sufficient performance).

**Physics-informed design of sustainable alloys considering real-world constraints**

Following the data–knowledge–application flow in **Fig. 1c**, we introduced a physics-informed ensemble (PIE) learning strategy (**Fig. 1d**) that integrates materials descriptors (**Supplementary Note S1**), multiscale physical principles, empirical rules, and semi-quantitative physical models (**Supplementary Note S2**). This strategy enables the PIE model to adhere to established physical laws and hypotheses governing alloy behavior across spatial and temporal scales [33-36]. First, we employed the PIE model to generate a digital alloy library with systematically varied compositions, microstructures, and properties—thus enabling continuous exploration across a broad CMP space. Second, the CMP features of these candidates were mapped in a standardized feature space to identify feasible processing routes from their experimentally explored nearest neighbors (**Fig. 1e**). This step establishes the detailed CMPP profiles for each candidate. Third, for sustainability-focused optimization, we quantitatively assess the operational expenditures (OPEX) and carbon footprint of each candidate based on its CMPP profile and feedstock type (e.g., primary metals or secondary scrap) (**Fig. 1f**). This step enables the joint optimization of feedstock selection, alloy composition and processing routes toward target performance and sustainability requirements.

**Development of a physics-informed ensemble model for Mg alloys**

To develop a PIE model for Mg alloys, we collected a Mg alloy benchmark dataset from published literature and our own experiments. The collected dataset comprises detailed CMPP profiles of Mg alloys (**Supplementary Table S1**), including chemical compositions (spanning 24 elements in reported Mg alloys), matrix microstructures (e.g., grain size), precipitate characteristics (e.g., type, size, and volume fraction), mechanical properties (e.g., yield strength (YS), tensile strength (TS), and elongation), and the manufacturing processes (e.g., cast, rolling, extrusion, forging, annealing, etc.). Mg alloys typically contain minor alloying additions—usually less than 20 wt. %—which are introduced to tailor their microstructure and mechanical properties. Mg alloys are generally classified as rare-earth (RE) Mg alloys containing elements such as Y, Gd, and Nd, or non-RE Mg alloys typically based on Al, Zn, Mn, and Ca. As shown in **Fig. 2a**, the collected Mg alloys are discretely distributed near the RE element and non-RE element corners. Notably, yield strength spans a wide range in both RE and non-RE Mg alloys, indicating that composition alone does not fully determine the mechanical properties.

We therefore collected cross-scale microstructural features of the Mg alloys along with their respective manufacturing process details (**Supplementary Table S1 and Table S2**).

As shown in **Fig. 2b**, the dataset comprises 52.8% RE-Mg alloys and 47.2% non-RE Mg alloys, each with chemical composition, grain size, mechanical properties, and manufacturing features. We noted, though, that information about precipitates was often incomplete: approximately 50% of entries lack data on precipitate type, 88% lack precipitate fraction, and 90% are missing size information. Indeed, many studies do not report precipitate features or use incomplete characterization to probe them. This shows the challenge of compiling complete and high-quality datasets, highlighting a characteristic limitation in alloy design, particularly when microstructure plays an important role. Thus, despite decades of research, the experimentally explored Mg alloy and microstructure space remains sparsely occupied.

To enable performance-oriented design of Mg alloys, we developed and compared a suite of models to quantify the CMP relationships learned from the benchmark dataset (see descriptors in **Supplementary Table S3**). As shown in **Fig. 2c**, we first established a classical physical model (**Supplementary Note S2**), including the Hall-Petch relation for inverse dependence of strength on grain size and solid-solution strengthening. We then evaluated a mathematical model using symbolic regression (**Supplementary Note S3**) and a conventional ML model via a decision tree algorithm (**Supplementary Table S4**). Although they achieve reasonable predictive performance, they rely on high-dimensional input data and are computationally costly. Moreover, their limited interpretability and intrinsic generalizability may yield unreliable predictions when applied to practical, 'real-world' alloy design, where processing and scale dependence matter.

To overcome these limitations, we developed the PIE model through a two-step ensemble strategy (**Extended data Fig. 1**). To embed physical laws into the ML model, we designed descriptors reflecting classical strengthening mechanisms and incorporated physics-based models alongside data-driven components, improving both interpretability and performance. To mitigate model bias, we trained a suite of single ML models (one-step) as baselines. We then selected the two best-performing models, random forest and XGBoost regressors (**Supplementary Fig. S1, Table S5a and Table S5b**), for the first-step ensemble via soft voting. During this process, $k$-fold cross-validation ($k$ = 5) was employed to generate multiple sub-models to conduct rigorous model evaluation, which were then integrated in the second-step ensemble to construct the final PIE model.

Given the incomplete microstructural information and the limited number of fully characterized entries, we constructed two variants of the PIE models. The first variant, the matrix-based PIE model, was trained using all available features except those describing precipitate characteristics (**Supplementary Table S6**). Building on this, we developed a second variant, an enhanced matrix- and precipitate-based PIE model, by integrating the top-ranked features (e.g., the ten most important features; see **Supplementary Note S5** for details of the feature screening strategy, **Table S7** for the descriptors, and **Table S8** for feature ranking) with precipitate-related features from the subset of samples where such data are available. This two-tier modelling strategy enables robust property prediction while effectively accommodating the data sparsity commonly encountered in real-world alloy development.

As shown in **Fig. 2c–2e**, the two-step PIE model achieves improvements in predictive performance compared to the one-step models while maintaining a balance between model complexity and computational efficiency (**Fig. 2f and Supplementary Note S6**). Since these models represent common ML approaches for many alloy and broader materials systems [37-39], our PIE model demonstrates superior benchmark performance on the Mg alloy family. In addition, SHAP (SHapley Additive exPlanations) analysis and invariant sensitivity analysis (see **Extended data Fig. 2 and Supplementary Fig. S2 and S3**) confirm that the PIE model captures the strengthening mechanisms and enables quantitative interpretation of composition and microstructural contributions to Mg alloy strength (also see the **Supplementary Fig. S4** for the transparent decision path of the tree-based model). The matrix-based PIE model is more robust across a wider range of Mg alloys because it was trained on a relatively large-scale dataset. In comparison, the matrix-and-precipitate-based PIE model achieves higher predictive accuracy for precipitate-containing alloys due to enhanced training with additional precipitate descriptors on a smaller, specialized dataset (see **Supplementary Fig. S5** for the statistical evaluation of predictions).

We then used the trained PIE model to generate a virtual Mg library by systematically exploring plausible combinations of composition—including principal alloying elements (e.g., Al, Zn, and RE elements) and tramp elements (e.g., Si, Fe, Cu, and Ti)—and microstructural features, and predicting the corresponding strengths. This yielded over $10^6$ unique Mg alloy entries with fully defined CMP features (**Supplementary Note S7**). As shown in **Fig. 2g**, the PIE-generated library spans a continuous compositional space, enabling exploration beyond experimentally sampled regions. The candidates also exhibit continuously distributed microstructural features (**Fig. 2h**), enabling microstructure-oriented exploration across the design space.

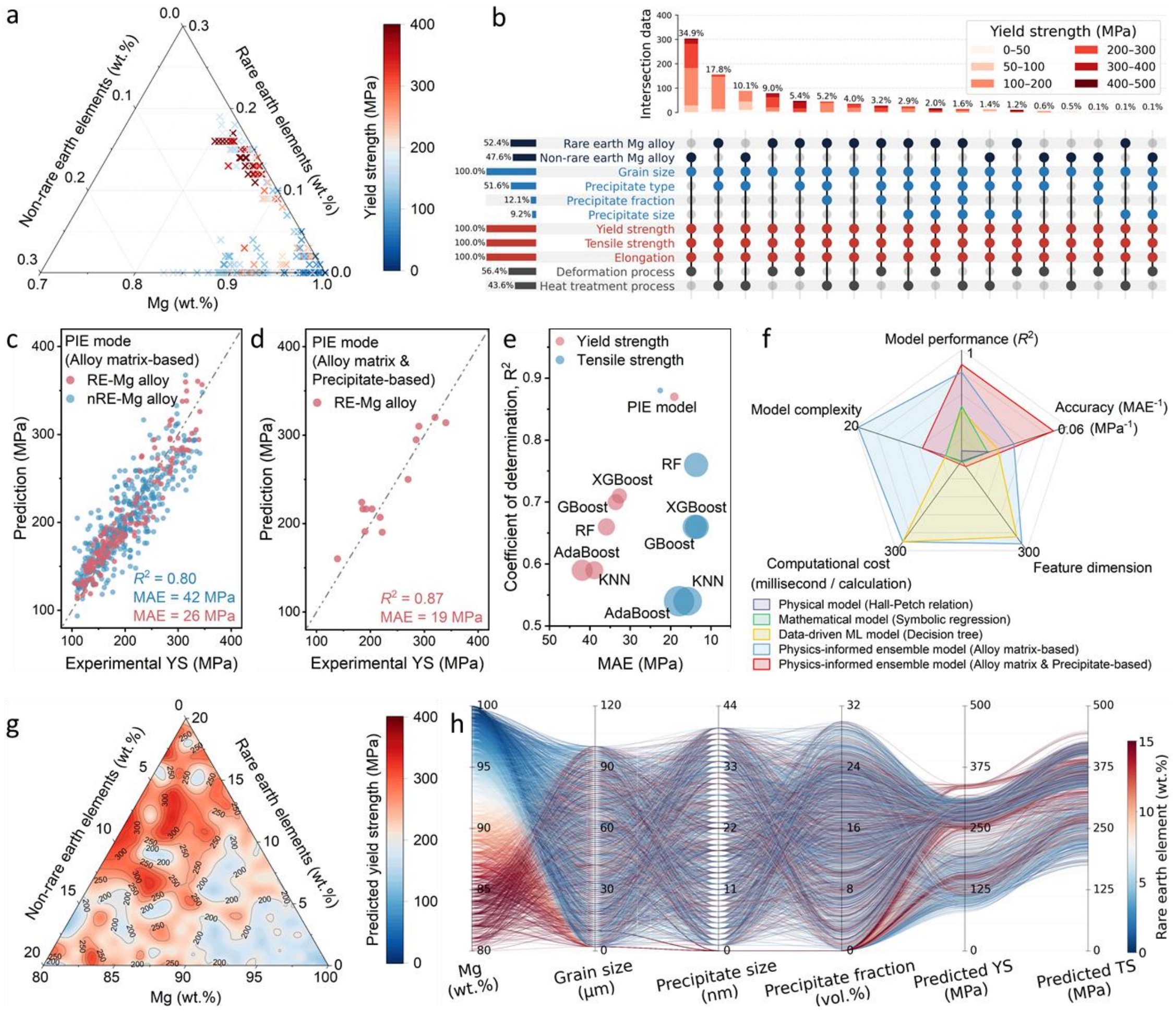

**Fig. 2 | Physics-informed ensemble learning and digital library of Mg alloys. Benchmark dataset (~$10^3$ entries): a**. Chemical composition distribution, colored by yield strength. **b**. Feature availability and co-occurrence, where columns represent the fraction of entries with specific feature combinations, and rows denote individual features; solid dots indicate included features. **Model performance: c**. PIE model trained using alloy matrix features. **d**. PIE model incorporating both matrix and precipitate features. **e**. Performance comparison of five common ML models (RF, KNN, GBoost, XGBoost, and AdaBoost) for yield strength and tensile strength; circle size represents mean absolute percentage error (larger values indicate higher error). **f**. Comparative evaluation of performance, complexity, feature dimensionality, and computational cost across physical, mathematical, black-box ML, and PIE models. **PIE-generated digital library (~$10^6$ candidates): g**. Chemical composition distribution, colored by predicted yield strength. **h**. Feature distribution of generated Mg alloys, including grain size, precipitate size, precipitate fraction, and predicted strength; line color indicates rare-earth element content.

**Transition to the real world: data-driven discovery of processing routes**

Designing viable industrial processing routes for a given alloy with defined CMP profiles remains highly challenging and marks one of the biggest obstacles when leveraging AI methods for real-world material development, due to the non-unique, path-dependent nature of the underlying CMPP relationships [40]. Traditionally, process design meanders between bounds set by thermodynamics, kinetic transients, costs, quality, and metallurgical intuition, with CMPP relationships established through long-term empirical and trial-and-error efforts [41]. Here, we propose a data-driven strategy that infers processing routes from target CMP features using K-nearest-neighbor (KNN) classification against reference alloys with known processing histories.

As illustrated in **Fig. 3a**, the typical industrial processing route for Mg alloys involves casting, homogenization annealing, deformation, and aging [41]. The specific processing steps and approaches depend on the alloy system (i.e., chemical composition) and the final product shape, both of which critically influence the resulting microstructure and properties. For instance, the chemical composition—particularly the presence or absence of RE elements—governs phase structure and formability, directly influencing the design of subsequent processing steps. Microstructural characteristics such as grain size are typically shaped by deformation, while precipitate fraction and size are further modified by heat treatment. The interplay between deformation and thermal treatment can introduce competing effects, as solid-solution strengthening and precipitate strengthening may counteract softening mechanisms [42]. Therefore, mechanical properties such as yield strength and tensile strength serve as indirect indicators of processing history, as they are intrinsically linked to composition and microstructural evolution.

To quantitatively evaluate the performance of KNN-based processing-route inference, we apply leave-one-out cross-validation (LOOCV). The corresponding confusion matrix (**Fig. 3b**) shows high classification accuracy across the five processing routes at a neighborhood size of $k = 3$, with diagonal values representing the class-specific correct classification rates. To further visualize the distribution of target alloys relative to the reference clusters, we apply *t*-distributed stochastic neighbor embedding (*t*-SNE) to project the CMP features into a two-dimensional latent space, where alloys with similar CMP characteristics are positioned close together. **Supplementary Fig. S6** illustrates the distribution of material features across the embedding space. As shown in **Fig. 3c**, RE-Mg and non-RE Mg alloys with different grain size exhibit a distinct distributions in the two-dimensional *t*-SNE map. This visualization is consistent with the KNN-LOOCV analysis, and further illustrates how specific target alloys are positioned relative to reference alloys with established processing histories. For instance, Mg-Al-Zn (AZ) alloys with fine grain size (~5 μm) cluster near alloys processed via hot deformation (upper right of **Fig. 3c**). Similarly, RE-Mg alloys with small grains and high strength are located near alloys processed through deformation followed by aging (lower left of **Fig. 3c**). Together with the KNN-inferred processing routes, we establish CMPP profiles for the PIE-generated Mg alloy candidates, providing a foundation for downstream sustainability assessment and interdisciplinary optimization.

**Sustainability assessment and application-oriented feature design**

Designing sustainable alloys requires balancing interdisciplinary criteria, including mechanical performance, carbon emissions, and manufacturing costs. Here, based on the CMPP profile of Mg alloys, we first quantify the operational expenditures (OPEX) from feedstock and processing costs, accounting for historical variations in feedstock and electricity prices and route-dependent consumption of processing consumables (e.g., protective gas $SF_6$ and $CO_2$) (**Supplementary Note S8**). Second, we estimate carbon emissions using feedstock-associated greenhouse-gas emission factors, accounting for variations associated with metal-production routes, energy mixes and life-cycle inventories (**Supplementary Note S8**). For secondary feedstocks, we account for their price relative to primary metals and adopt a cut-off approach for the emission assessment, excluding burdens from the previous product life cycle while accounting for secondary-materials recovery and preparation.

We first quantify the composition-dependent trends in OPEX and carbon emissions (**Fig. 3d and Fig. 3e**), and then incorporate processing routes and associated cost and emission ranges to map Mg alloy candidates with different strengths relative to as-cast pure Mg in the cost–emissions space (**Fig. 3f**). Notably, RE-Mg alloys exhibit high strength but tend to cause higher costs and carbon emissions, aligning with the concept of indirect sustainability. Non-RE Mg alloys are more cost-effective and environmentally friendly but typically exhibit lower strength. This result reveals the trade-off between cost, sustainability, and strength that must be resolved in real-world alloy design. We demonstrate a strategy to overcome this trade-off by incorporating secondary feedstock. As shown in the third quadrant of **Fig. 3f**, using recycled Mg and Al alloys to produce Mg-Al-X alloys provides a notable advantage by simultaneously reducing the OPEX and carbon emissions while maintaining strength, thus meeting the criteria for direct sustainability (lower emissions in production). This potential strategy balances Mg alloys among cost, carbon footprint, and strength, and can be tailored to region-specific geopolitical and resource considerations.

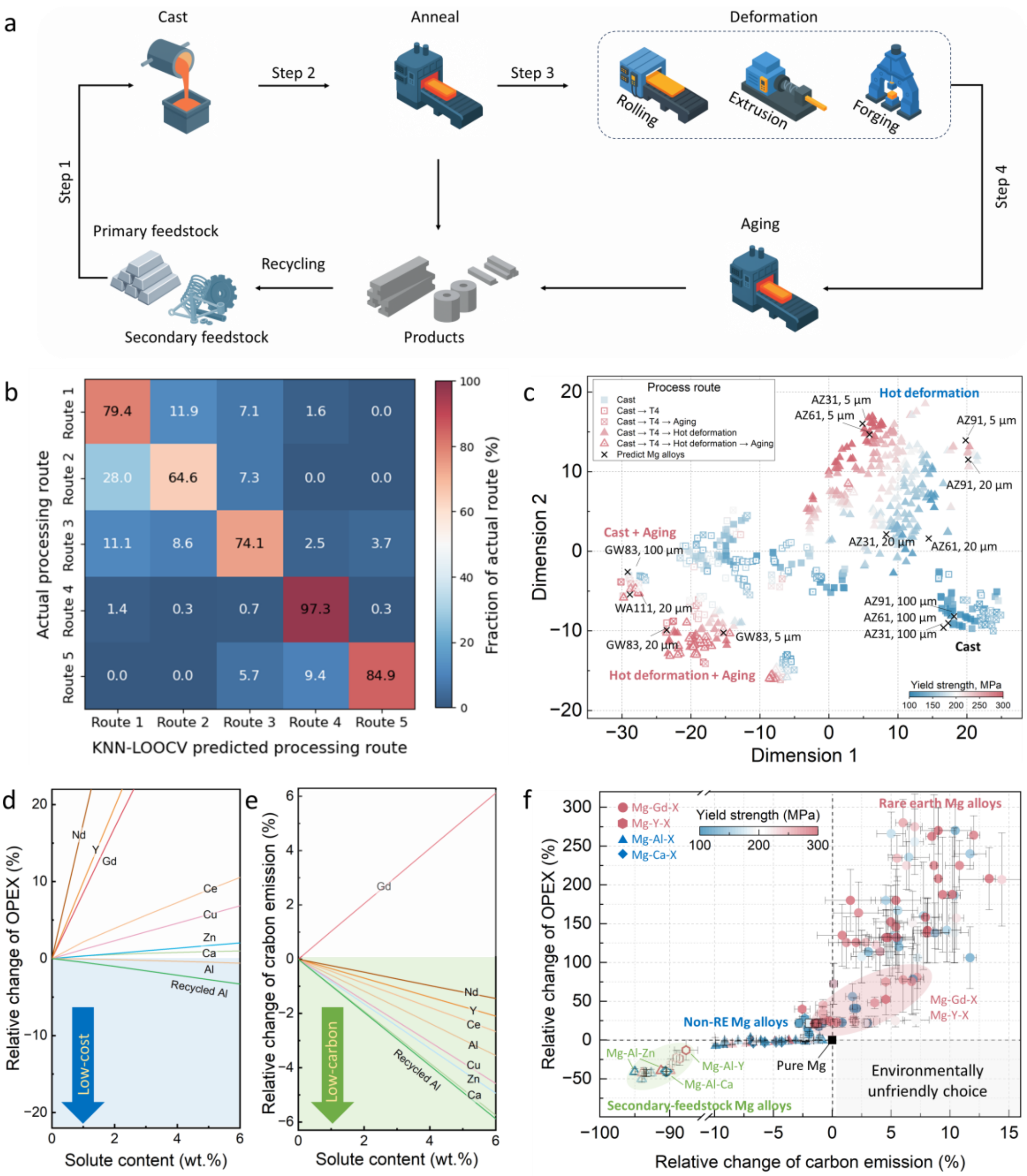

**Fig. 3 | Data-driven process mapping and sustainability assessment. a.** Schematic of typical industrial processing steps for Mg alloy manufacturing processes. **b.** Confusion matrix of KNN processing-route classification evaluated by LOOCV at neighbourhood size of $k = 3$. Values indicate the normalized proportion $N_{ij} / N_i$, where $N_{ij}$ is the number of samples from actual class $i$ predicted as class $j$ and $N_i$ is the total number of samples in class $i$. **c.** Two-dimensional *t*-SNE embedding of Mg alloys. Marker shape indicates process route; color denotes yield strength. **d.** Composition-dependent changes in operational expenditures (OPEX) and **e.** carbon emissions of as-cast Mg alloys, both relative to as-cast pure Mg. **f.** Distribution of Mg alloys in sustainability space, benchmarked against as-cast pure Mg. Error bars indicate arising from historical ranges in feedstock and electricity prices and feedstock-associated greenhouse-gas emission factors.

Guided by the sustainability landscape of Mg alloys (**Fig. 3f**), Mg alloys containing Gd generally exhibit superior yield strength, whereas substitution with Al offers more potential for reducing cost and carbon emissions. Therefore, we experimentally validated two representative Mg alloy series: (1) indirectly sustainable high-strength Mg-Gd-Y-Zn (GW) alloys (red circle in **Fig. 3f**), acting through weight-related property advantages; and (2) directly sustainable Mg-Al-Zn (AZ) alloys fabricated using recycled pure Mg scraps and recycled 6061 Al alloy (green circle in **Fig. 3f**), acting through better sustainability when produced. The alloy compositions, microstructural features, and predicted strength values are summarized in **Supplementary Table S9**. The process route for the Mg-8Gd-3Y-Zn (GW83) alloy (grain size: 5 μm) and the Mg-3Al-Zn (AZ31) alloy (grain size: 20 μm) was identified in **Fig. 3c**. Experimentally, the high-strength GW83 alloy was prepared following the processing route: Cast → T4 treatment → Forging → Aging (see *Methods* for detailed processing parameters). The chemical composition of the alloy was 7.6 wt.% Gd, 3.1 wt.% Y and 0.9 wt.% Zn (**Supplementary Fig. S7a**). The tensile stress-strain curves are shown in **Fig. 4a**. For the high-performance Mg alloy, it measures a yield strength of 296 ± 8 MPa, tensile strength of 398 ± 11 MPa, and elongation of 14 ± 1 %. The inverse pole figure (**Fig. 4b**) shows a refined recrystallized structure with a grain size of 4.8 μm (**Supplementary Fig. S8a**). A high-angle annular dark-field image and its corresponding diffraction pattern (**Fig. 4c**) confirm the presence of nanoscale precipitates. Notably, the Mg-Gd-Y-Zn alloy developed in this work achieves improvements of 72.8% in YS and 61.1% in TS over previously reported Mg-Gd-Y-based alloys within a comparable sustainability window (red-circled region in Fig. 3f).

For realizing a directly sustainable Mg alloy, we used pure Mg scrap and recycled 6061 Al scrap (composition provided in **Supplementary Table S10**) to minimize both environmental impact and production cost. According to the predicted results in **Supplementary Table S9**, the process route was as follows: Cast → T4 treatment → Extrusion (see ***Method*** for detailed processing parameters). The inverse pole figure (**Fig. 4d**) confirms a fully recrystallized grain structure, while the bright-field image (**Fig. 4e**) reveals a solid-solution matrix and is free of intragranular precipitates. The secondary-feedstock AZ31 Mg alloy contained 3.27 wt.% Al and 1.16 wt.% Zn (**Supplementary Fig. S7b**) and had a grain size of 14.5 μm (**Supplementary Fig. S8b**), achieved yield strength of 155 ± 4.2 MPa, tensile strength of 284 ± 1 MPa, and elongation of 16.8 ± 1.4 % (blue curves in **Fig. 4a**). Compared to primary-feedstock AZ31 alloy, the secondary-feedstock AZ31 alloy exhibits comparable mechanical performance while significantly reducing OPEX by 40.4% and carbon emission by 88.9%, **Fig. 4f**. The results demonstrate the practical effectiveness of the PIE framework in designing both high-performance and sustainable Mg alloys that meet real-world economic, geo-political, and environmental constraints.

In conclusion, the experimental validation of sustainable Mg alloys demonstrates a shift in alloy design philosophy, moving from a low-dimensional performance-centric and 'laboratory-only' paradigm toward a more holistic, 'real-world' design approach balancing optimization of performance, processing cost, and environmental impact. Our PIE framework provides an end-to-end solution that leverages cross-scale material descriptors with economic and environmental metrics, bridging materials science, processing engineering, and sustainability assessment within a unified modeling framework. By embedding physical knowledge, data-driven learning, and real-world manufacturing constraints into a unified framework, this work establishes a practical pathway for designing application-oriented, sustainability-aware alloys beyond empirical trial-and-error. More broadly, it highlights the potential of physics-informed, data-driven approaches to address emerging societal and industrial demands in materials development.

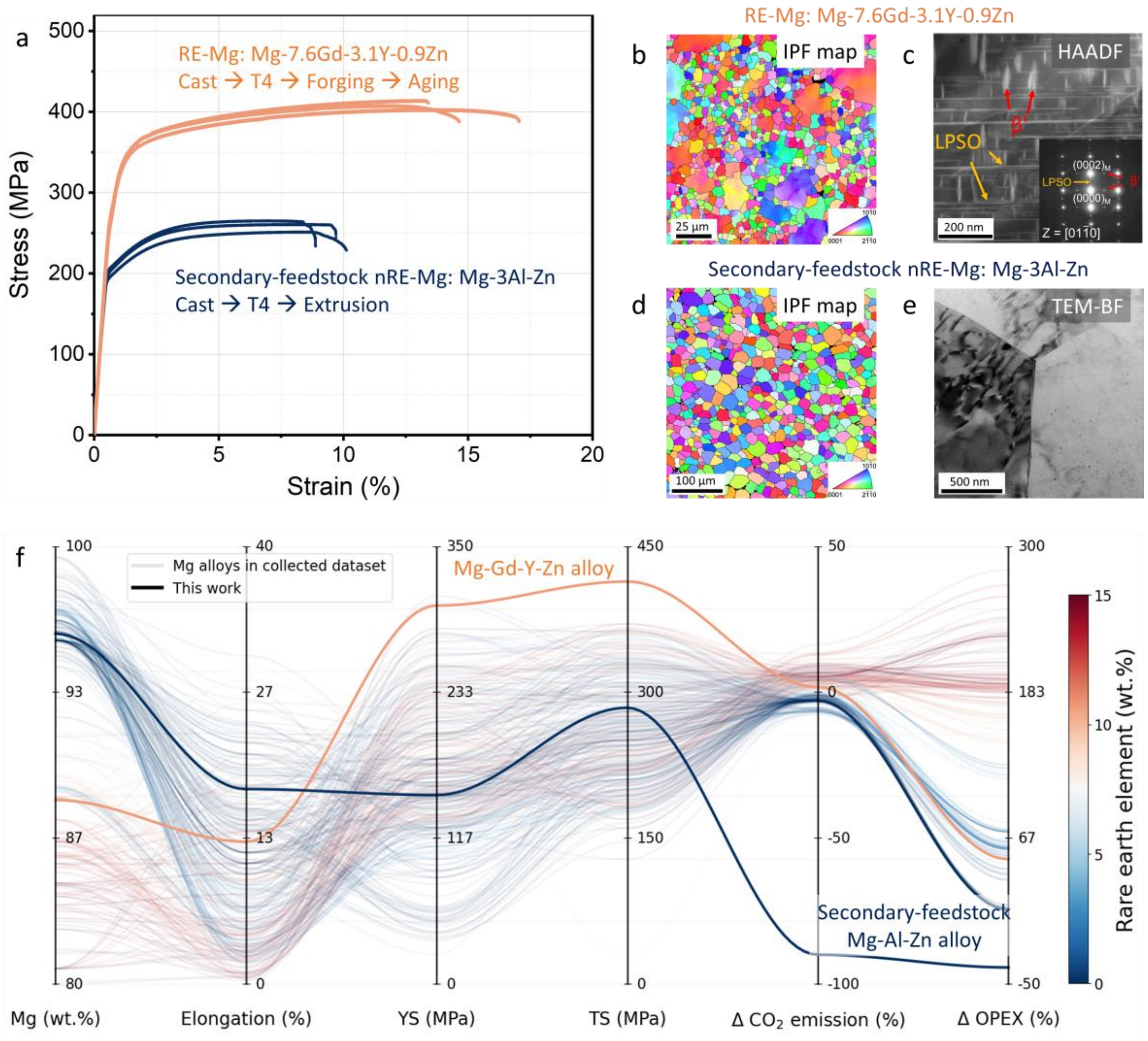

**Fig. 4 | Sustainable Mg alloys. a.** Tensile stress-strain curves of high-strength RE-Mg alloy (Mg-7.6Gd-3.1Y-0.9Zn alloy, orange curves) and secondary-feedstock nRE-Mg alloy (Mg-3Al-1Zn alloy, blue curves). **b** and **c**. The microstructure of the Mg-Gd-Y-Zn alloy, where the inserted diffraction pattern confirms the long-period stacking ordered (LPSO) phase and β' phase. **d** and **e**. The microstructure of the secondary-feedstock Mg-Al-Zn alloy, where the inserted diffraction pattern confirms the single Mg phase. **f**. Performance of Mg alloys under sustainability constraints. The Mg-Gd-Y-Zn alloy developed in this work (orange curve) is compared with previously reported Mg-Gd-Y-based alloys [43,44] within a sustainability window of 10% ⩽ ΔOPEX ⩽ 100% and -2% ⩽ $\Delta CO_2$ emission ⩽ 10%, relative to as-cast pure Mg. The secondary-feedstock Mg-Al-Zn

alloy (blue curve) is compared with primary-feedstock AZ31 alloys [45,46].

## Methods

### Mg alloy data collection and curation

The benchmark dataset of Mg alloys was manually collected from the literature and in-house experiments. A primary dataset comprising 827 Mg alloy entries was assembled, including chemical composition, microstructure, mechanical properties (yield strength, tensile strength, and elongation), and processing history (**Supplementary Table S1**). In addition, a subset of 50 Mg alloys was curated with detailed precipitate descriptors including distribution, morphology, habit plane, geometric size, and volume fraction.

### Feature engineering and descriptor design

To involve comprehensive information on the compositions of Mg alloys, we employed the materials informatics toolkits ***Matminer*** to expand the elemental features from 24 to 272 dimensions (also see **Supplementary Note S4** for the a detailed description of feature expansion). The compositional features combined with the microstructure and the estimated strength from empirical equations, resulting in a maximum of 277 alloy feature dimensions. To better incorporate the feature of microstructure, we classify the non-quantitative features (e.g., precipitate distribution) and directly use the numerical value for the quantitative features (e.g., grain size) (**Supplementary Table S2**).

To screen the most effective physical descriptors, we built a sequence of sub-models with different descriptors (**Supplementary Table S3**). We evaluated the comprehensive performance, such as coefficient of determination ($R^2$), mean absolute percentage error (MAPE), and mean absolute error (MAE), of each model with different descriptors (**Supplementary Note S1**). As a result, we determine the optimal descriptors for strength prediction with consideration of materials-based features and physical principles.

### Machine learning model training

The training process for the ensemble learning model consists of two steps: **(1) Pre-training of individual sub-models**, and **(2) Training the ensemble model using a soft voting strategy** (i.e., two-step voting).

In the first step, ten different sub-models were trained individually using the same optimal descriptors and train-test split. The models were trained using ten regression algorithms, including Gradient Boosting Regression (GBR), Random Forest Regression (RFR), XGBoost Regression (XGBR), Adaptive Boosting Regression (ADBR), K-Nearest Neighbors Regression (KNR), Support Vector Regression with linear kernel (SVR-linear), Decision Tree (DT), Ridge Regression (RR), Elastic Net Regression (ENR), and Lasso Regression (LR). Hyperparameters for each model were optimized via grid search and five-fold cross-validation, using 90% of the dataset. The remaining 10% was held out for final validation.

During cross-validation, the dataset was split with a 4:1 ratio between training and validation folds. As a result, each sub-model was trained on 70% of the total data in each fold. This approach ensured both sufficient training data and reliable performance evaluation. **Supplementary Table S4** summarizes the optimized hyperparameters of all sub-models, and **Supplementary Fig. S1** illustrates their predictive performance. Based on this evaluation, we identified the top 5 individual sub-models (GBR, RFR, XGBR,

KNN, and ADBR) for subsequent ensemble learning.

In the second step, we applied a two-step voting strategy (**model voting** and **dataset voting**) to train the ensemble model on the remaining 10% unseen data. For model voting, we evaluated eleven different weight combination strategies to identify the optimal ensemble model from the five sub-models. This strategy enables comparable and complementary predictive capabilities, enhancing the accuracy and robustness of the ensemble model while simultaneously limiting the computational complexity and cost of the voting process. In practice, we determined the optimal weights following three principles: (1) Performance-Oriented Weighting: Higher weights are assigned to models with lower MAE, emphasizing superior predictive performance; (2) Model Diversity Constraint: Models with structural differences (e.g., algorithm type of feature representation) or predictive differences (e.g., average output deviation) exceeding 0.2 are selected for combination, ensuring complementarity and reducing correlated errors; (3) Controlled weight variance: The variance of the assigned weights is constrained to less than 0.2 to prevent dominance by a single model and to enhance robustness. As a result, we identified the optimal sub-models and their corresponding weights: for yield strength prediction, XGBR was assigned a weight of 0.7 and RFR a weight of 0.3; for tensile strength prediction, both XGBR and RFR were assigned equal weights of 0.5. Furthermore, for dataset voting, we trained the ensemble models using five-fold cross-validation, and equal weights were assigned to models from each fold during the final voting stage. Ultimately, the PIE model was constructed by voting on 10 sub-models (5 folds × 2 sub-models). The voting strategy and corresponding performance metrics of ensemble models are summarized in **Supplementary Tables S5a-S5b** and **Table S6**.

We evaluate the performance of the model by using the $R^2$ and MAE:

$$R^2=1-\frac{\sum_{i=1}^{n}(y_i-\widehat{x_i})^2}{\sum_{i=1}^{n}(y_i-\overline{x_i})^2} \tag{1}$$

$$\mathrm{MAE}=\frac{\sum_{i=1}^{n}|y_i-\widehat{x_i}|}{n} \tag{2}$$

where $y_i$, $\widehat{x_i}$, and $\overline{x_i}$ are the experimentally measured value, predicted value, and average value of the experimentally measured values.

**Shapley additive explanations analysis**

Shapley Additive Explanations (SHAP) is a unified framework for interpreting machine learning predictions based on Shapley values from cooperative game theory. For a given model *f* and input *x*, the prediction is decomposed into contributions from each feature. The SHAP value $\emptyset_i$ for feature *i* is defined as[47, 48]:

$$\emptyset_i=\sum_{S\subseteq N\{i\}}\frac{|S|!(|N|-|S|-1)!}{|N|!}[f_x(S\cup\{i\})-f_x(S)] \tag{3}$$

where N is the set of all input features, S is a subset of features not containing *i*, and $f_x(S)$ is the expected model output when only features in *S* are known. The final model output is approximated as:

$$f(x)=\emptyset_0+\sum_{i=1}^{M}\emptyset_i \quad (4)$$

where $\emptyset_0$ is the base value (e.g., the expected model output), and $\emptyset_i$ represents the contribution of feature *i* to the prediction. SHAP values provide consistent, locally accurate, and model-agnostic explanations of both global feature importance and individual predictions.

**Invariant sensitivity analysis**

Invariant sensitivity analysis was used to evaluate the impact of individual features on model predictions, enabling interpretation and understanding of the model's decision-making process[49]. For a well-trained model, invariant sensitivity analysis involves two main steps: (1) selecting one feature while keeping all other features fixed, and (2) predicting the model output as the selected feature varies, then plotting this variation against the predicted results to generate a sensitivity curve. Discontinuities or sharp changes in the sensitivity curve indicate the turning point where the selected feature significantly influences the model's predictions. In this work, we performed invariant sensitivity analysis on key features of matrix- and precipitate-based PIE model, including the *d*-electron number, grain size, and precipitate fraction.

**K-nearest-neighbor classification**

K-nearest-neighbor (KNN) classification was used to quantitatively infer processing routes from the composition–microstructure–property (CMP) descriptors of Mg alloys. The input descriptors included the major alloying elements (Al, Zn, Mn, Gd and Y), microstructural descriptors (grain size and fractions of *β* precipitates and LPSO phases), and mechanical properties (YS and TS). All descriptors were standardized before classification to account for differences in feature scales. A total of 633 samples with known processing histories were grouped into five representative processing-route classes: 'casting'; 'casting → T4 treatment'; 'casting → T4 treatment → aging'; 'casting → T4 treatment → hot deformation'; and 'casting → T4 treatment → hot deformation → aging'. KNN classification was performed in the standardized CMP feature space, with the processing route assigned according to the majority class among the *k* nearest neighbors (**Supplementary Fig. S9 and Table S12**). A neighborhood size of $k = 3$ was selected for subsequent processing-route classification.

**Leave-one-out cross-validation**

Leave-one-out cross-validation (LOOCV) was used to evaluate the performance of the KNN processing-route classification. In each iteration, one of the 633 samples was withheld as the test sample and its processing route was predicted from the remaining samples using KNN; this procedure was repeated until each sample had served once as the test sample. Classification performance was quantified using overall accuracy ($N_{correct}$ / $N_{total}$), balanced accuracy ($\frac{1}{C}\sum_{i=1}^{C}\frac{TP_i}{TP_i+FN_i}$), precision ($P_i=\frac{TP_i}{TP_i+FP_i}$), recall ($R_i=\frac{TP_i}{TP_i+FN_i}$), and F1-score ($F1_i = \frac{P_i R_i}{P_i + Rl_i}$), where $N_{correct}$ and $N_{total}$ denote the numbers of correctly classified and total samples, respectively; $C$ is the number of processing-route classes; and $TP_i$, $FP_i$, and $FN_i$ denote the numbers of true-positive, false-positive, and false-negative predictions for class $i$, respectively. Overall accuracy represents the fraction of correctly classified samples, whereas balanced accuracy represents the mean recall across the five processing-route classes, thereby accounting for class imbalance. Precision, recall, and F1-score were calculated for each processing-route class, together with sample-weighted and class-balanced averages. A confusion matrix was constructed from the LOOCV predictions to quantify class-specific correct classifications and misclassifications among the five processing routes.

**_t_-distributed stochastic neighbor embedding (_t_-SNE)**

*t*-distributed stochastic neighbor embedding (*t*-SNE) was used to visualize the distribution Mg alloys according to their composition-microstructure-property (CMP) characteric [50]. To ensure consistency with the KNN processing-route classification, *t*-SNE used the same standardized CMP descriptors, including the major alloying elements (Al, Zn, Mn, Gd and Y), microstructural descriptors (grain size and fractions of β precipitates and LPSO phases), and mechanical properties (YS and TS). All descriptors were standardized to zero mean and unit variance before dimensionality reduction.. A perplexity of 50, learning rate of 100, maximum of 1,000 iterations and random state of 42 were used. In the resulting two-dimensional map, marker shapes indicate the known processing routes of the reference alloys.

**Quantitative sustainability metrics assessment**

Operational expenditures (OPEX) and carbon emissions were evaluated based on alloy composition, processing route and feedstock source. The OPEX accounts for feedstock costs and processing-related costs associated with electricity and consumables during melting, deformation and heat treatment. Carbon emissions were estimated from feedstock-associated greenhouse-gas emission factors and processing energy consumption. Variations in feedstock and electricity prices and feedstock-associated greenhouse-gas emission factors were considered to assess the robustness of the calculated OPEX and carbon emissions. For secondary feedstocks, price variations relative to primary metals were considered, while a cut-off approach was adopted for the emissions assessment, excluding burdens from the previous product life cycle while accounting for secondary-material recovery and preparation. Detailed model definitions, calculations, parameters, data sources and system boundaries are provided in **Supplementary Note S8**.

**Preparation of Mg alloys**

The Mg-8Gd-3Y-Zn alloy was prepared by high-purity Mg (⩾ 99.9%), purity Zn (⩾ 99.9%), Mg-25Y (wt.%), and Mg-30Gd (wt.%) master alloys. The Mg-3Al-1Zn (AZ31) alloy was prepared by pure Mg scraps (⩾ 99.9%) and 6061 Al alloys. The alloys were melted under the protective gas of $CO_2$ and $SF_6$ at 750 °C, followed by isothermally holding for 30 min before casting. The ingots of alloys were solution treated at 500 °C for 5 h, followed by water quenching to room temperature. The solution-treated Mg-8Gd-3Y-Zn alloy was deformed by multi-directional forging at 390 °C with axial deformation of 50%. Subsequently, the forged Mg-8Gd-3Y-Zn alloy was annealed at 225 °C for 60 h and then cooled in the air. The solution-treated AZ31 alloy was hot-extruded at 350 °C with an extrusion ratio of 20:1, followed by air cooling.

**Tensile test**

The tensile test was conducted according to the GB/T228.1-2010 standard. Cylindrical specimens were prepared with a gauge radius of 5 mm and a gauge length of 25 mm. The strain rate for the tensile test was set to $10^{-3}$ $s^{-1}$. Independent tensile tests were conducted three times for each alloy to assess the consistency of mechanical performance. The yield strength (YS) of the Mg alloy was determined as the 0.2% offset strength. The tensile strength (TS) was defined as the maximum strength of the Mg alloy before fracture.

**Microstructure characterization**

Bulk specimens for electron back-scattered diffraction (EBSD) and energy-dispersive X-ray Spectroscopy (EDS) characterization were prepared by electro-polishing in an ethanol solution with 10 vol.% perchloric acid, followed by ion-beam polishing using Fischione SEM-Mill-1060 operated at 1 keV for 0.5 h. Thin film specimens for transmission electron microscopy (TEM) characterization were cut into 3 mm discs and ground to a thickness of 50 μm, followed by ion-beam milling using Gatan PIPS-695 operated at 6 keV and 3 keV.

EBSD characterization was conducted on Zeiss Sigma-500 SEM with a step size of 0.5 μm at 15 kV. TEM bright-field (BF) images and high-angle annular dark-field (HAADF) images were captured by a JEOL-2100F TEM at 200 kV. The corresponding EDS mapping was collected by Oxford Instruments during the collection of EBSD and HAADF images. The description of composition and microstructural features of Mg alloys related to experimental characterization techniques is listed in **Supplementary Table S11**. To quantify precipitation features, we conducted image contrast analysis on the SEM-BSE image and the TEM-HAADF image. For microscale precipitates, the length and width are defined as the length along the major and minor axes of the precipitate. For nanoscale precipitates, the length, width, and thickness are well-defined by the crystallographic relation between the precipitate and Mg matrix. Detailed criteria for experimental determination of precipitate size can be found in **Supplementary Note S9**.

**Data availability**

The database of training data and generated Mg alloys is available in the Supplementary Information and https://www.aimslab.cn/#/query_alloy.

**Code availability**

The implementations of the codes used for calculating the features and training the models are publicly available at https://github.com/nwpusl/AI_Alloy. The codes of SHAP analysis are available at https://github.com/slundberg/shap.

**Competing interests**

The authors have obtained a Software Copyright in China for the physics-informed ensemble framework described in this work (Registration No. 2024R11S1265852). The authors declare no other competing financial or non-financial interests.

**Acknowledgments**

The authors are grateful for the financial support provided by the National Natural Science Foundation of China (Grant Nos. 52425101, 52471012, 52127801, 32301040, 52401216), the Advanced Materials-National Science and Technology Major Project (2024ZD0607400) and the Shanghai Science and Technology Project (Nos.24010700100 and 23JC1402300), and the funding by HORIZON-European Innovation Council (EIC)-2022-PATHFINDEROPEN-Project-ID 101099736-CoCoMag. Q.L. acknowledges the support from the Outstanding Doctoral Graduates Development Scholarship of Shanghai Jiao Tong University. M. J.-K. gratefully acknowledges financial support from the MPG-FhG cooperation project MaRS.

**Author contributions**

X.Z., J.L., and D.R. conceived the idea. Q.L. and Z.W. contributed equally to this work. Q.L. led the experimental investigation and conducted the formal analysis. Z.W. led the algorithm development. Q.L., B.F., and J.W. conducted the experiments. Q.L., C.L., J.W., and W.Z. collected the experimental dataset. Z.W. and L.S. developed the machine learning model and the digital alloy library. Q.L., Z.W., L.S., M. J.-K., Z.R., J.L., D.R., and X.Z., discussed and interpreted the experimental results and machine learning models. X.Z. and J.L. supervised the research and secured funding. Q.L. and Z.W. wrote the original paper. All authors discussed the results in detail and contributed to the writing, revision, and editing of the submitted manuscript.

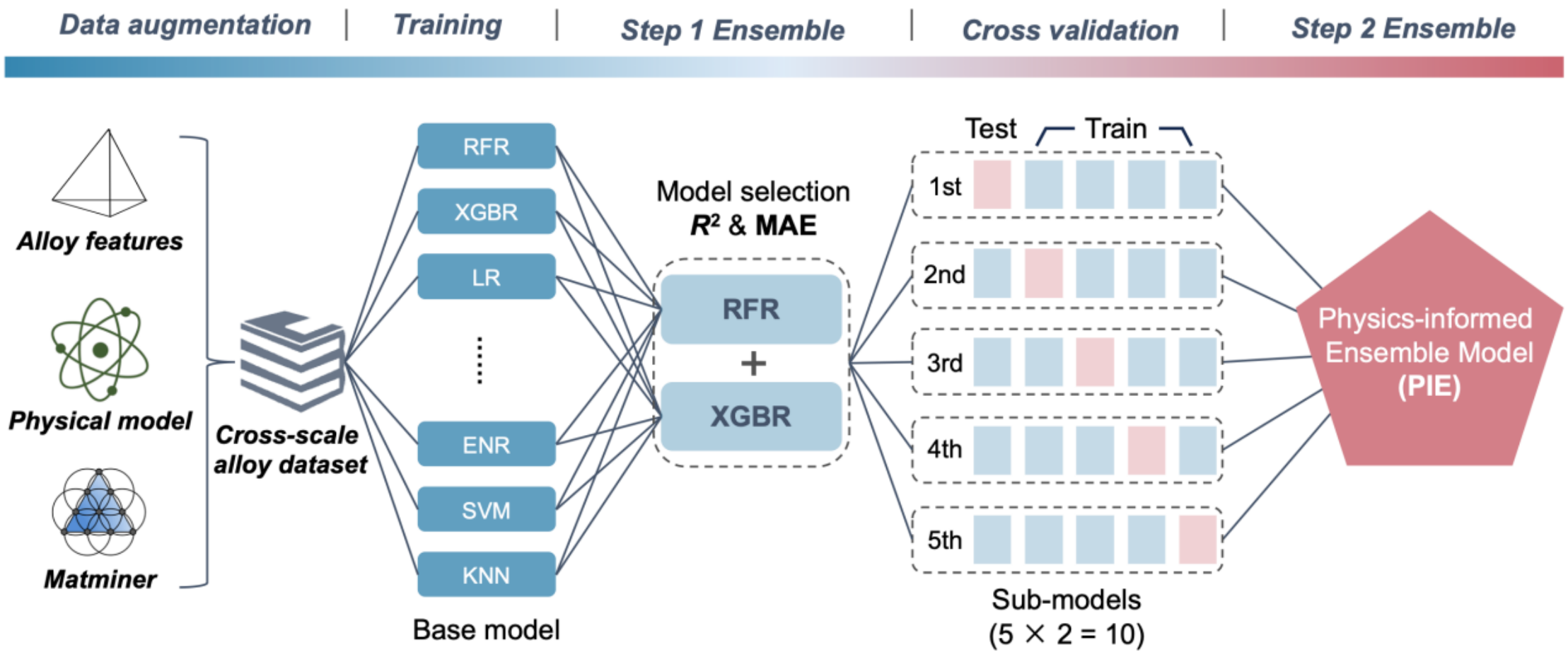

**Extended Data Fig. 1 | Two-step ensemble strategy for the development of the PIE model**

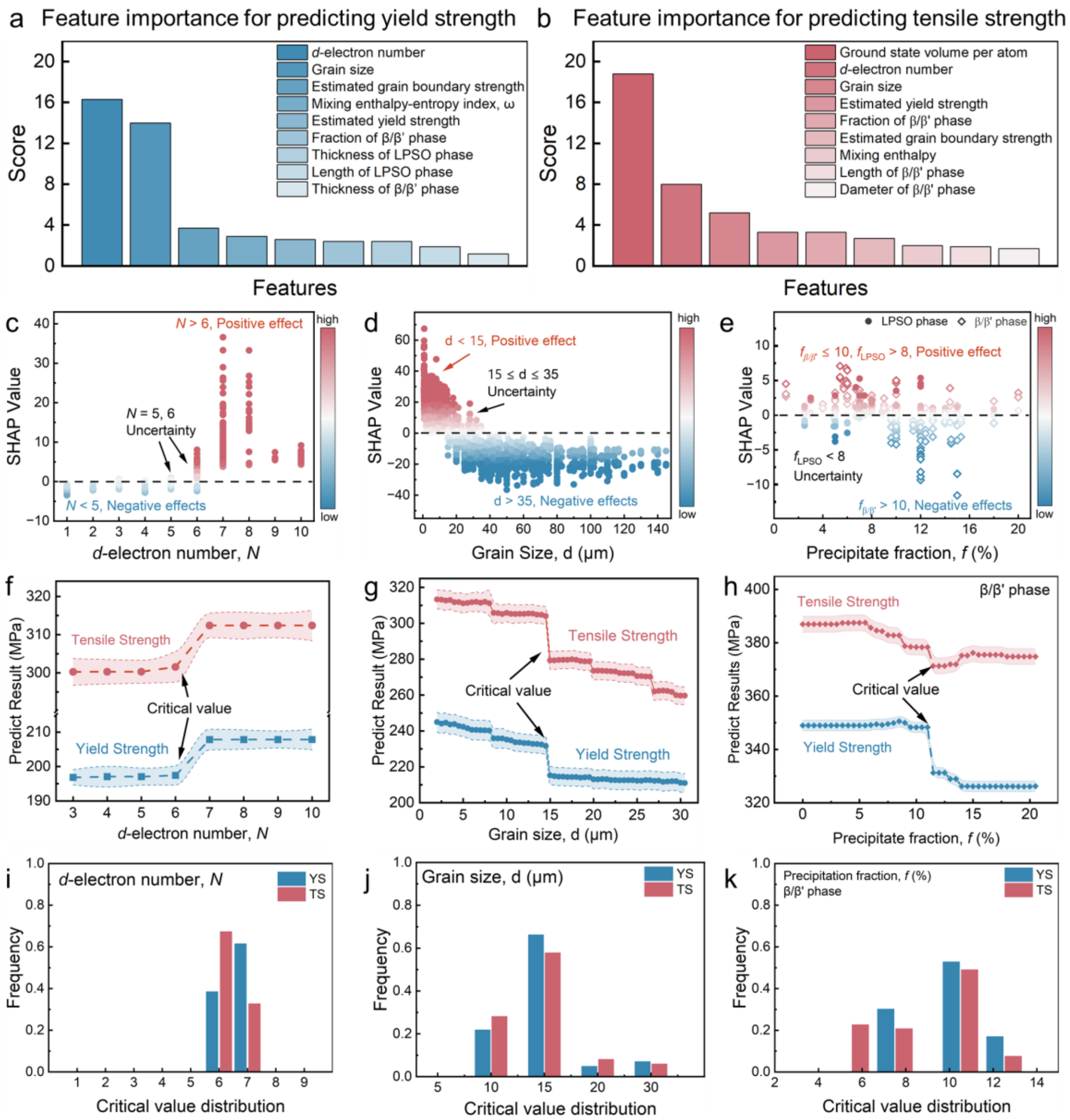

**Extended Data Fig. 2 | Physical explanation of the physics-informed model.** Feature importance rankings (**Supplementary Table S8**) for matrix-and-precipitate-based PIE model in prediction (a) yield strength and (b) tensile strength. SHAP value distributions for key features: (c) *d*-electron number, (d) grain size, and (e) precipitate fraction. Feature values with exclusively positive SHAP contributions (red) indicate a positive effect, whereas those with exclusively negative contributions (blue) indicate a negative effect; values exhibiting both are classified as uncertain (grey). The transition point at which the SHAP effect shifts from negative to positive is defined as the critical value. Invariant sensitivity analysis confirms critical thresholds for (f) *d*-electron number, (g) grain size, and (h) β/β' phase fraction. Statistical distributions of critical values across alloys for key features: (i) *d*-electron number, (j) grain size, and (k) β/β' phase fraction.